\documentclass[twocolumn,showkeys,prd,nofootinbib,floatfix,preprintnumbers]{revtex4-1}

\usepackage[utf8]{inputenc}
\usepackage{multirow}
\usepackage{amsfonts,amsmath,amssymb} 
\usepackage{graphicx,graphics,color}
\usepackage{gensymb}
\usepackage{longtable}
\usepackage{subfigure}
\usepackage{amsmath}

\usepackage{hyperref}
\usepackage[normalem]{ulem}
\hypersetup{
    colorlinks=true,
    linkcolor=blue}

  \def\apjl{ApJ}

\newcommand{\be}{\begin{equation}}
\newcommand{\ee}{\end{equation}}
\newcommand{\ba}{\begin{eqnarray}}
\newcommand{\ea}{\end{eqnarray}}

\begin{document}

\title{Constraining Self-interacting Scalar Field Dark Matter From the Black Hole Shadow of the Event Horizon Telescope}

\author{Gabriel G\'omez}
\email{gabriel.gomez.d@usach.cl}
\affiliation{Departamento de F\'isica, Universidad de Santiago de Chile,\\Avenida V\'ictor Jara 3493, Estaci\'on Central, 9170124, Santiago, Chile}

\author{Patrick Valageas}
\email{patrick.valageas@ipht.fr}
\affiliation{Université  Paris-Saclay,  CNRS,  CEA,  Institut  de  physique  théorique,  91191,  Gif-sur-Yvette,  France}

\begin{abstract}

An exciting possibility to constrain dark matter (DM) scenarios is to search for their gravitational 
imprints on Black Hole (BH) observations. In this paper, we investigate the impact of self-interacting 
scalar field DM on the shadow radius of a Schwarzschild BH. We implement a self-consistent formulation, 
paying attention to the enhancement of the DM density due to the BH gravitational influence and 
the accretion flow onto the BH.
First, we calculate the first-order correction to the shadow radius caused by a general DM environment. 
Then, we apply this perturbative method to the case of self-interacting scalar field DM and derive 
analytical expressions for the critical impact parameter. 
We find that self-consistency requirements, involving the lifetime and the mass of the central 
DM soliton, or the mass and the size of the extended virialized DM halo, ensure that the impact
of the DM environment on the shadow radius is below the observational upper bound.
This emphasizes the importance of taking into account the self-consistency constraints of the
underlying DM scenario, which can strongly limit the range of possible DM density profiles and
their impact on the shadow radius.

\end{abstract}

\maketitle

\section{Introduction}


An intriguing prediction in general relativity (GR) is the existence of BHs \cite{Poisson:2009pwt,thorne2000gravitation}. Apart from the physical conceptions encoded in the spacetime metric, BHs are an ideal laboratory to study high-energy astrophysical phenomena that take place in their strong field regimes. Furthermore, the presence of non-trivial (e.g. scalar, vector) field profiles, or fields that are non-minimally coupled to gravity, surrounding BHs may modify the spacetime structure, leading to long lived hairs \cite{Herdeiro:2015waa,Cardoso:2016ryw,Herdeiro:2014goa}. Hence, violations of the no-hair theorem can provide evidence of new physics \cite{Cardoso:2011xi,Barranco:2011eyw,Cardoso:2016ryw,Herdeiro:2014goa}. These fundamental concerns have motivated different astrophysical experiments, such as  the Even Horizon Telescope (EHT) \cite{EventHorizonTelescope:2019dse,2022ApJ...930L..17E,EventHorizonTelescope:2022wkp}, GRAVITY collaboration \cite{GRAVITY:2020gka} and the LIGO and VIRGO collaborations \cite{Abbott:2016blz,LIGOScientific:2017ync}, among the most important scientific projects, to test the properties of BHs. So far all the data are consistent with GR, but cannot rule out the possibility of other non-trivial metrics within the current uncertainties (see e.g \cite{Vagnozzi:2022moj}).  

While convincing evidences for the existence of BHs have finally been obtained through, among other observational inferences,  the images of BHs Messier (M) 87$^{\star}$ and Sagittarius (Sgr) A$^{\star}$ shadows,  many fundamental questions in physics are still under scrutiny. One intriguing problem  is related to the nature of the still elusive DM, which makes up around 80$\%$ of the matter content of the Universe \cite{Planck:2018vyg}. 
The standard Cold DM (CDM) model is able to reproduce most observational data at both cosmological and galactic scales. However, it faces some tensions on small scales, such as 
the inner structure of dwarf spheroidal and low-mass spiral galaxies \cite{Oh:2010mc}, 
the core-cusp \cite{Moore:1994yx,deBlok:2009sp} and the too-big-to-fail problems \cite{BoylanKolchin:2011de,Garrison-Kimmel:2014vqa}. Moreover, CDM particles such as
weakly interacting massive particles have not been detected yet.
Scalar field (SF) models are other well-motivated candidates that arise in extensions of the standard model of particle physics to describe the DM component of the Universe. Thus, DM can be in the form of axion-like particles \cite{Turner:1983he,Sikivie:2009qn,Marsh:2010wq}, ultra light bosonic fields  (Fuzzy DM)  \cite{Hu:2000ke,Hui:2016ltb} or self-interacting SF called also Bose Einstein Condensate (BEC) DM \cite{Goodman:2000tg,Boehmer:2007um,Lee:2008jp,Harko:2011xw,Suarez:2013iw,Fan:2016rda,Brax:2019fzb}.  

An attractive feature of SFs is that they can condensate into macroscopic structures (with macroscopic occupancy) called \textit{soliton cores}, providing a possible explanation for the observed DM cores in some galactic halos. Several works have studied how the gravitational influence of a central supermassive BH can affect such DM solitons, pointing out that the soliton core can survive cosmological time scales \cite{UrenaLopez:2002du,Barranco:2011eyw,Barranco:2012qs,Avilez:2017jql}. Thus, long-lived self-gravitating configurations could reside at the center of galaxies in company with BHs on time scales relevant for astrophysical observations. In the case of self-interacting SF, the soliton core is formed by the balance between self-gravity and a repulsive self-interaction.  The free case corresponds to a class of Fuzzy DM models where the soliton core is supported by
the so-called quantum pressure (due to the wavelike dynamics associated with the
Schr\"odinger equation) instead of the self-interactions.  The dynamics and phenomenological consequences of ultralight SFs surrounding a BH have been studied extensively in the literature, employing both  Newtonian and relativistic approaches \cite{Cruz-Osorio:2010nua,Barranco:2011eyw,Ferreira:2017pth,Bar:2018acw,Bar:2019pnz,Hui:2019aqm,Desjacques:2019zhf,Davies:2019wgi} (see also references therein). However, more limited attention has been given to the self-interacting SF scenario. For instance, Brax et al. \cite{Brax:2019npi} employed an analytical relativistic treatment to study the infall of the SF onto the central Schwarzschild BH (see also Ref.~\cite{Ravanal:2023ytp} for the case of charged BHs), while Chavanis et al. used a Gaussian ansatz \cite{Chavanis:2019bnu} and velocity dispersion tracing \cite{Chavanis:2019amr} as part of their analytical approach. In this paper, following \cite{Brax:2019npi}, we consider the repulsive self-interacting model, as an alternative to Fuzzy DM scenarios.   

An exciting possibility for observing a smoking gun for DM identification is to search for gravitational interactions with BHs and simultaneously probe possible deviations from the Schwarzschild (and Kerr) metric due to a DM environment. 
Interestingly, the enhancement of the DM density caused by BHs may lead to potentially detectable signatures in observations of gravitational waves emitted by BH binaries \cite{Eda:2013gg,Kavanagh:2020cfn,Boudon:2023vzl}, the movement of S2 stars around the Galactic center \cite{Bar:2019pnz,Becerra-Vergara:2021gmx} and, of main interest in this paper, the shadow size of BHs inferred by the observed ring-like images of M87$^\star$ and Sgr A$^\star$ by the EHT (see e.g. \cite{Lacroix:2012nz} for an early work). All these observations are unprecedented examples of  high-precision measurements of the gravitational field of BHs in the strong-field regime.  The latter of the mentioned observational programs has been the subject of intense research, after the release of the first M87$^\star$ EHT results, due to the possibility of addressing fundamental problems in physics \cite{Vagnozzi:2022moj,2022ApJ...930L..17E}.  Thus, DM can affect in different ways BH observations whereby a self-consistent modeling of the spacetime geometry where BH and DM coexist is crucial.   

It is important to mention that most of the existing works describing the spherically symmetric spacetime metric around a BH immersed in a DM halo have used a Newtonian treatment to calculate the metric coefficients from the tangential velocity and a given DM mass distribution \cite{Xu:2020jpv,Xu:2021dkv}. Consequently,  the effect of the BH on the DM distribution has not always been taken into account in a self-consistent manner when computing some BH astrophysical properties \cite{Pantig:2022toh,Zhang:2021bdr,Saurabh:2020zqg,Anjum:2023axh,Liu:2021xfb}.  However, \cite{Lacroix:2012nz} considered the spike CDM profile, which arises from the adiabatic growth of BHs, to constrain the properties of the DM distribution from the angular size of the BH shadow. On the other hand, \cite{Daghigh:2022pcr} used the the Tolman-Oppenheimer-Volkoff equations to construct the spacetime metric describing a  BH surrounded by  a DM spike. Recently, \cite{Figueiredo:2023gas} implemented a numerical treatment to study the effect of a generic DM profile with accretion growth on geodesic properties.  In the context of SFs, \cite{Pantig:2022sjb}  studied the effect of the soliton core in the fuzzy DM model (or wave DM) on a supermassive BH.  They used empirical data for the shadow diameter provided by EHT to constrain the DM properties. However, the self-interacting SF model has not been consistently studied yet. Some works have considered the standard core halo to describe the DM distribution around BHs \cite{Xu:2020jpv,Xu:2021dkv}. Building on the self-consistent modeling developed 
in \cite{Brax:2019npi}, we present a theoretical approach that incorporates two key aspects overlooked previously:  i) the effect of the BH gravitational potential on the DM soliton core, and ii) the modification of the metric due to the modified DM soliton core. We use a perturbative treatment that takes into account the DM accretion and self-gravity. From this formulation, we calculate the perturbed shadow radius associated with a soliton core as well as with the outer virialized DM halo.
We find that self-consistency requirements, taking into account the loss of matter to the central
BH and the lifetime of the central DM soliton, imply that the impact of the DM environment onto
the shadow radius is below the observational upper bound.

This paper is organized as follows. In Section~\ref{sec:Perturbative}, we present the perturbative 
scheme used to derive the dark matter contribution to the metric and to the BH shadow radius. 
This treatment is very generic and does not assume a specific dark matter model. 
In Section~\ref{sec:theoretical}, we provide an overview of the SF dark matter model
and of the scalar field profile at different radii. 
In Section~\ref{sec:impact-steady} we apply our perturbative scheme to the SF dark matter model
and present our results for the shadow radius, examining the impact of both the central soliton
and the extended virialized halo. 
Finally, we discuss the main findings of this work in Section~\ref{sec:discussion}. 

\section{Impact of the environment on the metric and BH shadow radius}
\label{sec:Perturbative}

The results of this section are applicable to generic spherically symmetric DM and/or 
baryonic distributions around a Schwarzschild BH.

\subsection{Perturbative scheme}
\label{sec:Perturbative-scheme}

In this paper we focus on spherically symmetric systems, that is,
a Schwarzschild BH at the center of a spherical DM cloud.
Then, using Schwarzschild coordinates, the line element can be expressed as
\begin{equation}
ds^{2}=-f(r)dt^{2}+g(r)dr^{2}+r^{2}d\Omega^{2},
\end{equation}
where $f(r)$ and $g(r)$ are the metric coefficients, $r$ is the radial coordinate and $d\Omega$ is the solid angle. 
As we consider a finite-size scalar cloud, at large distance we must recover the
vacuum Minkowski metric, $f \to 1$ and $g \to 1$ for $r \to \infty$.
Defining as usual the enclosed mass $m(r)$ by
\begin{equation}
g(r) = \left( 1 - \frac{2 {\cal G} m(r)}{r} \right)^{-1} ,
\end{equation}
the Einstein tensor reads as
\begin{align}
G^t_t & = - \frac{2 {\cal G}}{r^2} \frac{\partial m}{\partial r} , \label{eqnGtt} \\
G^r_t & = \frac{2 {\cal G}}{r^2} \frac{\partial m}{\partial t} , \label{eqnGrt} \\
G^r_r & = - \frac{2 {\cal G} m}{r^3} + \left( 1 - \frac{2 {\cal G} m}{r} \right) \frac{1}{r f} \frac{\partial f}{\partial r} .
\label{eqnGrr}
\end{align}

We use a first-order perturbative scheme 
to obtain the metric perturbation due to the DM environment,
taking advantage of the fact that at all radii beyond the BH
horizon we can perform a perturbative expansion around the Schwarzschild metric.
This is because close to the BH the gravitational field is dominated by the BH and the metric is close to the Schwarzschild metric,
whereas at large radii, where the DM self-gravity comes into play,
we are already far in the weak gravity regime and the metric is
close to the Minkowski metric, which in this regime is also close to the Schwarzschild metric.
This perturbative approach would break down at the horizon, where
$f=0$ at zeroth order, but we only need to consider radii beyond the
photon sphere, which is greater than the Schwarzschild radius by a factor $3/2$. Therefore, this perturbative scheme is well defined and sufficient for our purposes.

Thus, at zeroth order we consider the Schwarzschild metric, which is the solution of the Einstein
equations in vacuum, around a spherically symmetric mass $M_0$.
This gives the well-known expressions
\begin{equation}
m_0(r)= M_0 , \;\;\; g_0(r) = \frac{1}{1-\frac{2{\cal G} M_0}{r}} , \;\;\; 
f_0(r) =  1-\frac{2{\cal G} M_0}{r} .
\end{equation}
Next, to include the effect of a generic DM distribution, we write the metric functions as
\begin{equation}
f = f_0 + \delta f , \;\;\; g = g_0 + \delta g , \;\;\; m = m_0 + \delta m , \;\;\; 
\delta g = \frac{2 {\cal G} \delta m}{r f_0^2} ,
\end{equation}
and we work at linear order over all the perturbations. 
We restrict our computations to radii above a radius $r_{\rm min}$ greater than the BH horizon, so that $f_0$ and $g_0$ are finite and nonzero and the perturbative scheme is well defined.
Then, the Einstein equations read $\delta G^\mu_\nu = 8\pi {\cal G} T^\mu_\nu$, where $T^\mu_\nu$ is the energy-momentum tensor of the DM (or more generally of the environment).
This gives
\begin{align}
& \frac{\partial \delta m}{\partial r} = - 4\pi r^2 T^{t}_{t},\label{eqntt} \\
& \frac{\partial \delta m}{\partial t} = 4\pi r^2 T^{r}_{t},\label{eqnrt} \\
& - \frac{2 {\cal G} \delta m}{r^2 f_0} + \frac{\partial \delta f}{\partial r} - \frac{2 {\cal G} M_0 \delta f}{r^2 f_0}  
= 8 \pi r {\cal G} T^r_r . \label{eqnrr}
\end{align}

As we consider a finite-size cloud of radius $R_{\rm cloud}$, at radii $r>R_{\rm cloud}$ we must recover
the vacuum Schwarzschild solution, albeit with a shifted mass $M_0+\delta M_0$ because of the
dark matter mass. This implies
\begin{equation}
r \geq R_{\rm cloud} \! : \;\; \delta f = - \frac{2 {\cal G} \delta M_0}{r} , \;\; \delta m = \delta M_0 , \;\;
\delta g = \frac{2 {\cal G} \delta M_0}{r f_0^2} .
\label{eq:f-m-g-Rsol}
\end{equation}
Assuming a steady state has been reached, the conservation equation 
$\nabla_\mu T^\mu_t=0$ leads to the condition of constant inward flux $F$,
\begin{equation}
\sqrt{fg} r^2 T_t^r = F .
\label{eq:flux-F}
\end{equation}
Using $f_0 g_0=1$ we obtain
\begin{equation}
\frac{\partial \delta m}{\partial r} = 4\pi r^2 \rho , \;\;\;
\frac{\partial \delta m}{\partial t} = 4\pi F ,
\end{equation}
where we defined $\rho = - T^t_t$.
This gives
\begin{equation}
\delta m(r) = m_{\min} + 4 \pi F t  + 4 \pi  \int_{r_{\min}}^{r} dr \, r^2 \rho , \label{eqndelta-m}
\end{equation}
where $m_{\min}$ is an integration constant. 
In the right-hand side, the second and third terms correspond to the mass that has fallen into the
BH since time $t$ and to the remaining mass in the scalar cloud.
Thus, the increase with time of the second term is balanced by the decrease of the third term.
In practice, we can choose to work at the time $t=0$.
The constant $m_{\min}$ corresponds to the dark matter mass inclosed below the arbitrary radius
$r_{\min}$.
Using the boundary condition (\ref{eq:f-m-g-Rsol}) at $R_{\rm cloud}$,
we can also write Eq.(\ref{eqndelta-m}) as
\begin{equation}
r \geq r_{\min} : \;\;\; \delta m(r) = \delta M_0 - 4 \pi \int_{r}^{R_{\rm cloud}} dr \, r^2 \rho . 
\label{eqndeltaM}
\end{equation}
For $r \! \geq \! R_{\rm cloud}$ we recover the Schwarzschild solution (\ref{eq:f-m-g-Rsol}).

Using the boundary condition $\delta f \to 0$ for $r \to \infty$, in agreement with (\ref{eq:f-m-g-Rsol}),
the Einstein equation (\ref{eqnrr}) can be integrated and gives
\begin{equation}
r \geq r_{\min} : \;\;\; \delta f = - f_0 \int_r^\infty dr \frac{r}{f_0} \left[ \frac{2 {\cal G} \delta m}{f_0 r^3} 
+  8 \pi {\cal G} P \right] ,
\label{eq:delta-f}
\end{equation}
where we defined the effective pressure $P=T^r_r$.
Using $\delta m = \delta M_0$ and $P=0$ for $r> R_{\rm cloud}$, we can check that we recover
the Schwarzschild solution (\ref{eq:f-m-g-Rsol}) for $r>R_{\rm cloud}$.

As expected, we can see that the results (\ref{eqndeltaM}) and (\ref{eq:delta-f}) do not depend on the
radius $r_{\min}$ introduced at an intermediate step of the computation. They are valid at all radii
$r$ sufficiently greater than $r_s$, so that $f_0$ and $g_0$ are always of order unity and different from zero.

\subsection{Perturbed Black Hole Shadow by Dark Matter}

One of the most fascinating properties of BHs is the \textit{shadow}: a dark region surrounded by circular orbits of photons known as the \textit{photon sphere}. This bright region is caused by gravitational light bending and photon capture at distances close to the even horizon.
We first consider the radius $r_{\rm ph}$ of the ``photon sphere'', associated with the unstable
circular orbits of light rays around the BH.
This radius is given by the implicit equation
\begin{equation}
r_{\rm ph} = 2 f \left( \frac{df}{dr} \right)^{-1} \bigg\rvert_{r=r_{\rm ph}}\label{rphoton} .
\end{equation}
The radius of the BH shadow is then the minimal impact parameter of photons escaping from the BH. Thus, the boundary of the shadow viewed by a distant observer is determined by the closest approach of photons before being captured, and it is tangent to the spherical orbits of photons close to the horizon \cite{Perlick:2021aok}. Photons with smaller impact parameters will eventually cross the horizon and fall onto the singularity.  This critical impact parameter can be computed in terms of the photon sphere as \cite{Psaltis:2007rv}
\begin{equation}
b_{\rm cr} = \frac{r_{\rm ph}}{\sqrt{f(r_{\rm ph})}} \label{bcr} .
\end{equation}
Then, the shadow angular radius $\alpha_{\rm sh}$ measured by a distant observer,
at $r \gg R_{\rm halo}$, is
\begin{equation}
\alpha_{\rm sh} = \frac{b_{\rm cr}}{r} .
\end{equation}
Thus, the BH shadow is determined by the critical impact parameter $b_{\rm cr}$.
As in Sec.~\ref{sec:Perturbative-scheme}, we compute the photon sphere radius and the critical impact parameter
at first order over the dark matter perturbation.
At zeroth order, we recover the standard result for a Schwarzschild metric
\begin{equation}
r_{\rm ph 0} =  3 {\cal G} M_0 = \frac{3}{2} r_s , \;\;\; r_s = 2 {\cal G} M_0 ,
\end{equation}
where $r_s$ is the Schwarzschild radius of the BH.
At first order, we obtain the deviation
\begin{equation}
\delta r_{\rm ph} = \frac{r_{\rm ph 0}^3}{6 {\cal G} M_0} \left[ \frac{\partial \delta f}{\partial r}
(r_{\rm ph 0}) - \frac{2 {\cal G} M_0}{r_{\rm ph 0}^2 f_0(r_{\rm ph 0})} \delta f (r_{\rm ph 0}) \right] .
\end{equation}
Using Eq.(\ref{eqnrr}), this simplifies as
\begin{equation}
\delta r_{\rm ph} = r_{\rm ph 0} \frac{\delta m(r_{\rm ph 0})}{M_0} + 4\pi {\cal G}  r_{\rm ph 0}^3 
P(r_{\rm ph 0})  .
\end{equation}
As expected, using Eq.(\ref{eqndelta-m}) with $r_s < r_{\rm min} < r_{\rm ph 0}$,
we can see that the location of the photon sphere
only depends on the metric functions and the dark matter distribution at radii below $r_{\rm ph}$.
The term $\delta m(r_{\rm ph 0})$ is simply the enclosed mass within the radius 
$r_{\rm ph 0}$. This is the generic effect due to the shift of the mass within
$r_{\rm ph 0}$ caused by the presence of dark matter.
The second term $P(r_{\rm ph 0})$, which would be zero for pressureless dust, is an additional
relativistic contribution that depends on the equation of state of the matter.

For the critical impact parameter, we obtain up to first order
\begin{equation}
b_{\rm cr 0} = 3 \sqrt{3} {\cal G} M_0 , \;\;\; \delta b_{\rm cr} = - \frac{9 \sqrt{3}}{2} {\cal G} M_0 
\delta f (r_{\rm ph 0})  ,
\label{eq:delta-b-cr}
\end{equation}
that is,
\begin{equation} 
    b_{\rm cr} = 3 \sqrt{3} {\cal G} M_0 \left(1 - \frac{3}{2} \delta f (r_{\rm ph 0})\right).
 \label{eq:delta-b-cr-2}   
\end{equation}
Notice that this result, along with Eq.~(\ref{eq:delta-f}), is independent of the DM or baryonic properties of the environment. It only assumes a Schwarzschild BH as the vacuum solution at zeroth order and a spherically symmetric DM and/or baryonic finite-size cloud.

\subsection{Non-degeneracy with a Schwarzschild BH}

In the computation described in the previous sections, we denoted $M_0$ the mass of the BH
without the DM cloud. However, in practice it may not be easy to measure $M_0$, as it can be
contaminated by the dark matter close to the horizon.
To be more explicit, let us assume that from the orbital dynamics of stars at a radius 
$R_{\rm dyn}$ one can measure the total enclosed dynamical mass 
\begin{equation}
M_{\rm dyn} = M_0 + \delta M_{\rm dyn} , 
\label{eq:M-dyn}
\end{equation}
with
\begin{equation}
\delta M_{\rm dyn} = \delta M_0 - 4\pi \int_{R_{\rm dyn}}^{R_{\rm cloud}} dr \, r^2  \rho .
\label{eq:deltaM-dyn}
\end{equation}
This dynamical mass measures the total mass located in the central region, that is, the sum
of the BH and dark matter, which determines the gravitational potential felt by neighbouring stars.
If $R_{\rm dyn}>R_{\rm cloud}$, $\delta M_{\rm dyn} = \delta M_0$ is the total mass of the dark matter
cloud, but if $R_{\rm dyn}<R_{\rm cloud}$, $\delta M_{\rm dyn} < \delta M_0$ is the fraction of dark matter
that is within the orbital radius of the observed stars.
Then, one can compare this mass $M_{\rm dyn}$ with the mass $M_{\rm sh}$ that would be
measured from the shadow angle, assuming a Schwarzschild BH in vacuum
\begin{equation}
M_{\rm sh} = \frac{b_{\rm cr}}{3 \sqrt{3} {\cal G}} .
\end{equation}
If these two measured quantities are equal, one can conclude that the system is probably made of 
a Schwarzschild BH of mass $M_{\rm dyn} = M_{\rm sh}$, or a degenerate system 
that obeys the same relation $M_{\rm dyn} = M_{\rm sh}$, such as a BH and a cloud of dust
of size smaller than the photon sphere.
In contrast, if one measures $M_{\rm dyn} \neq M_{\rm sh}$, one can conclude that the system
is not a Schwarzschild BH in vacuum.
Therefore, to constrain a dark matter cloud such as the one investigated in this paper,
we must consider the difference between these two observational definitions of the system mass.

Assuming we measure the dynamical mass (\ref{eq:M-dyn}), the predicted critical impact 
parameter would be
\begin{equation}
b_{\rm cr}^{\rm dyn} = 3 \sqrt{3} {\cal G} ( M_0 + \delta M_{\rm dyn}) ,
\end{equation}
if we interpret the data as arising from a Schwarzschild BH in vacuum.
This must be compared with the result (\ref{eq:delta-b-cr-2}).
Therefore, the departure from the isolated BH hypothesis is measured by the difference
\begin{eqnarray}
&& \Delta b_{\rm cr} = b_{\rm cr} - b_{\rm cr}^{\rm dyn} \nonumber \\
&& = b_{\rm cr 0} \left\{ \int_{r_{\rm ph 0}}^{R_{\rm cloud}} dr \left[ \frac{ {\cal G} ( \delta m - \delta M_0)}
{(r-r_s)^2} + \frac{4 \pi {\cal G} r^2 P}{r-r_s} \right ] \right. \nonumber \\
&& \left. + \frac{4\pi}{M_0} \int_{R_{\rm dyn}}^{R_{\rm cloud}} dr \, r^2 \rho \right\} ,
\label{eq:Delta-bcrit}
\end{eqnarray}
where we used Eq.(\ref{eq:delta-f}).
This quantity is non-zero when there is some scalar mass beyond $r_{\rm ph 0}$ 
($\delta m \neq \delta M_0$ at some radii in the integration range) and some dark matter pressure,
or there is some dark matter beyond the stellar orbits.
As expected, this probes the dark matter distribution between the photon sphere and the radius of the
DM cloud, which is the origin of the possible mismatch between $M_{\rm dyn}$ and $M_{\rm sh}$.

\subsection{Observations}

From the observational side, the EHT collaboration, a global very long baseline interferometer array, has imaged for the first time the central BH at the heart of the elliptical galaxy M87 \cite{EventHorizonTelescope:2019dse,Akiyama:2019bqs}. This earth-sized telescope has the capability of resolving the central compact radio sources as an asymmetric bright emission ring.  The observational data released in 2019 at the event-horizon scale image is consistent with theoretical predictions of GR for the shadow of the Kerr BH.  These unprecedented observations were followed by the image of Sgr A$^\star$ reported in 2022  \cite{2022ApJ...930L..17E,EventHorizonTelescope:2022wkp}, with a bright ring also consistent with a Kerr BH geometry.  They calibrated the geometrical BH shadow and the observed size of the ring images.  In particular, the EHT collaboration reported values of the fractional deviation, $\delta$, for the Sgr A$^{\star}$ BH. This quantity measures any deviation between the inferred shadow diameter and that of a Schwarzschild BH of angular size $\theta_{g}= \mathcal{G}M/Dc^{2}$ (in dimensional units). That is
\begin{equation}
    \delta = \frac{r_{\rm sh}}{r_{\rm sh,Sch}}-1,
\end{equation}
where $r_{\rm sh,Sch} = 3\sqrt{3} \theta_{g}$. They used prior information on the mass to distance ratio of the Sgr A$^\star$ BH based on dynamical analyses of the orbit of the Galactic center star S0-2, resulting in a posterior distribution with a small discrepancy of $\sim4\%$ for $\theta_{g}=5.125\pm0.009\pm0.020$~$\mu$as (VLTI) and $\theta_{g}=4.92\pm0.03\pm0.01$~$\mu$as
(Keck). See Ref.~\cite{EventHorizonTelescope:2022xqj} for more details. These inferences, in turn, translate into the following constraints for the fractional deviation
\begin{equation}
       \text{Keck}: \delta=0.04^{+0.09}_{-0.10},
\end{equation}
\begin{equation}
     \text{VLTI}: \delta=0.08^{+0.09}_{-0.09}.
\end{equation}
The goal of this paper is to use these observational values to constrain the properties of a possible scalar DM cloud around this supermassive BH. Thus, we will compare the theoretical prediction of such a SFDM cloud, as given by Eq.~(\ref{eq:Delta-bcrit}), with the above observational limits, which we can summarize as
\begin{equation}
\left| \frac{\Delta b_{\rm cr}}{b_{\rm cr}} \right|_{\rm obs} \lesssim 0.1 .
\label{eq:observ}
\end{equation}
The mass of the central BH, or overdensity, is estimated around
\be
M_{{\rm Sgr A}^\star} \simeq 4 \times 10^6 M_\odot, \;\;\; 
r_s \simeq 4 \times 10^{-7} {\rm pc} .
\label{eq:M-SgrA*}
\ee
This is obtained from the orbit of the S0-2 star, with a semimajor axis
\be
R_{\rm dyn} \simeq 5 \times 10^{-3} {\rm pc} .
\label{eq:Rdyn-SgrA*}
\ee

\section{Scalar field dark matter }\label{sec:theoretical}

\subsection{Enhancement of the soliton core by the BH}

Scalar field dark matter scenarios predict the formation of hydrostatic
equilibrium configurations, often called solitons, with a flat density core.
These spherically symmetric equilibria are also called boson stars when their radius
is comparable to astrophysical scales.
In the case of Fuzzy Dark Matter scenarios, with a scalar field mass $m_\phi \sim 10^{-22}$ eV,
such solitons have a radius $R_{\rm sol}$ of the order of 1 kpc and could partially explain
the small-scale tensions of the CDM model, such as the observation of flat galactic cores
instead of the cuspy density NFW profile found in dark matter-only numerical simulations\footnote{Such observations may also be explained by baryonic feedback effects. See e.g. \cite{DelPopolo:2016emo,Dutton:2018nop}.}.
In this model, these solitons are supported by the so-called quantum pressure and the
soliton radius is set by the de Broglie wavelength $\lambda_{\rm dB} = 2\pi/(m_\phi v)$,
where $v$ is the virial velocity of this DM halo.

In this paper, we focus instead on models with significant self interactions,
where the solitons are governed by the balance between gravity and the effective pressure
due to the self-interactions. We do not assume a priori specific values for the
DM parameters and do not require these solitons to reach kpc sizes.
We will investigate which general constraints can be derived from the observation of the
Sgr A$^\star$ shadow radius on the possible presence of such a DM soliton, or more generally
a DM halo, around the central Galactic supermassive BH.

The soliton DM density profile is affected by the BH at small radii, where the BH gravitational potential becomes more important than the DM self-gravity.
This leads to a DM spike below some transition radius, whereas the soliton profile remains
unchanged at larger radii. Various approaches have been implemented  to study this problem for the free case, mainly using numerical methods \cite{Clough:2019jpm,Bamber:2020bpu}. However, few works  have been devoted to study  the self-interacting case, with the exception of   \cite{Brax:2019npi}, which investigated the infall of a self-interacting SF onto the central Schwarzschild BH. 
We follow their approach, focusing on the large scalar mass regime $m_\phi \gg 10^{-17}$ eV
where their results apply, which we briefly recall below.

\subsection{Equations of motion}
\label{sec:motion}

We consider scenarios where the DM corresponds to a real scalar field $\phi$,
with the relativistic action \cite{Fan:2016rda}
\begin{equation}
\mathcal{S}_{\phi}=\int d^{4}x \sqrt{-g}   \left[-\frac{1}{2} g^{\mu\nu}\partial_{\mu}\phi \partial_{\nu}\phi
-V(\phi)\right],
\end{equation}
where $g_{\mu\nu}$ is the metric tensor and 
\begin{equation}
V(\phi)= m_\phi^2 \frac{\phi^2}{2} + \lambda \frac{\phi^4}{4}
\label{eq:quartic}
\end{equation}
is the SF potential with a quartic self-interaction coupling $\lambda$. Here $m_\phi$ is the DM particle mass and we take $\lambda>0$, which corresponds to a repulsive self-interaction
that can balance gravity and give rise to hydrostatic equilibria.
This defines the characteristic density $\rho_a$ and radius $r_a$ given by
\begin{equation}
\rho_a =  \frac{4 m_\phi^4}{3\lambda} , \;\;\; r_a = \frac{1}{\sqrt{4\pi{\cal G} \rho_a}} .
\label{eq:rho-a-def}
\end{equation}
The components of the energy-momentum tensor associated with the SF $\phi$ read as
\begin{align}
T^t_t &= -\frac{1}{2f} \left( \frac{\partial\phi}{\partial t} \right)^2
- \frac{1}{2g} \left( \frac{\partial\phi}{\partial r} \right)^2 - V(\phi), \label{eqnTtt} \\
T^r_t &= \frac{1}{g} \frac{\partial\phi}{\partial r} \frac{\partial\phi}{\partial t} , \label{eqTrt} \\
T^r_r &= \frac{1}{2f} \left( \frac{\partial\phi}{\partial t} \right)^2
+ \frac{1}{2g} \left( \frac{\partial\phi}{\partial r} \right)^2 - V(\phi),\label{eqnTrr} 
\end{align}
and the scalar field obeys the nonlinear Klein Gordon equation of motion
\begin{equation}
\frac{\partial^2\phi}{\partial t^2} - \sqrt{\frac{f}{g}} \frac{1}{r^2} \frac{\partial}{\partial r}
\left[ \sqrt{\frac{f}{g}}r^2\frac{\partial\phi}{\partial r} \right] + f m_\phi^2 \phi + f \lambda \phi^3 = 0.\label{eqnKG}
\end{equation}

In this paper, we focus on the large scalar-mass limit, 
\begin{equation}
m_\phi r_s \gg 1 .
\label{eq:large-mass}
\end{equation}
From Eq.(\ref{eq:M-SgrA*}) this corresponds to
\be
m_\phi \gg 10^{-17} \, {\rm eV} .
\label{eq:m_phi-min}
\ee
Thus the Compton wavelength of the scalar field is smaller than the BH horizon and
we have $\partial_r \ll m_\phi$. 
Then, at leading order the solution of the nonlinear Klein Gordon equation (\ref{eqnKG}) reads
\cite{Frasca:2009bc,Brax:2019npi}
\begin{equation}
\phi(r,t) = \phi_0(r) \, {\rm cn}[ \omega(r) t - {\bf K}(r) \beta(r), k(r) ],\label{eqnsolprofile}
\end{equation}
where ${\rm cn}(u,k)$ is the Jacobi elliptic function of argument $u$ and modulus $k$,
${\bf K}(k) = \int_0^{\pi/2} d\theta /\sqrt{1-k^2\sin^2\theta}$  is the complete elliptic integral of the first kind, 
$\phi_{0}$ is the amplitude of the oscillations and $\omega$ is the angular frequency of this anharmonic 
oscillator.  
We recover the harmonic oscillator when $k=0$ as ${\rm cn}(u,0)=\cos(u)$.
Thus, ${\rm cn}(u,k)$ allows us to describe the anharmonic oscillator associated with a cubic
nonlinearity, as in Eq.(\ref{eqnKG}) or in the standard Duffing equation.

The angular frequencies $\omega(r)$ at different radii are related by
\begin{equation}
\omega(r) = \frac{2 {\bf K}(r)}{\pi} \omega_0 ,
\label{eq:omega0-def}
\end{equation}
where $\omega_0$ is constant.
On the other hand, at leading order in the large scalar-mass limit, the temporal and radial derivatives 
of the SF profile, Eq.~(\ref{eqnsolprofile}), read as
\begin{align}
\frac{\partial\phi}{\partial t} &= \phi_0 \omega \frac{\partial {\rm cn}}{\partial u},\label{eqntemporal}\\
\frac{\partial\phi}{\partial r} &= - \phi_0 {\bf K} \beta' \frac{\partial {\rm cn}}{\partial u} + \dots,\label{eqnradial}
\end{align}
where the dots represent subleading terms. 
Substituting Eq.(\ref{eqnsolprofile}) into the Klein-Gordon equation Eq.(\ref{eqnKG}),
one obtains \cite{Brax:2019npi}
\begin{align}
\frac{\pi^2 f}{4 g} \beta'^2 &= \omega_0^2 - \frac{f m_\phi^2 \pi^2}{(1-2k^2) 4 {\bf K}^2},\label{eqnvel}\\
\frac{\lambda_4 \phi_0^2}{m_\phi^2} &= \frac{2k^2}{1-2k^2}.\label{eqnpar}
\end{align}
From the above expressions, the components of the energy-momentum tensor can be computed.

\subsection{Transonic scalar field profile}

\subsubsection{Transonic solution}
\label{sec:transonic}

In the steady state, we recover the integrated form of the continuity equation
(\ref{eq:flux-F}), where in the left-hand side we replace $T_t^r$ by its average 
$\langle T_t^r \rangle$ over the fast scalar-field oscillations.
Then, as for the classical Bondi problem of the spherical accretion of a gaseous cloud,
we obtain a unique transonic solution with a critical value of the accretion rate
\cite{Brax:2019npi},
\begin{equation}
F_c = F_\star \frac{r_s^2 m_\phi^4}{\lambda} = F_\star \frac{3}{4} \rho_a r_s^2 ,
\label{eq:Fc}
\end{equation}
with $F_\star \simeq 0.66$.
This result simply means that close to the horizon we are in the relativistic and
nonlinear regime, where the dark matter density is of the order of $\rho_a$ and the
velocity is of the order of the speed of light.
In contrast with the standard Bondi transonic solution for polytropic gases with an adiabatic
index $4/3 < \gamma < 5/3$, the sonic point is not much above the Schwarzschild radius.
This is because in the Newtonian regime the quartic self-interaction (\ref{eq:quartic})
leads to an effective pressure with a stiff equation of state, $P\propto \rho^\gamma$
with $\gamma=2$.

\subsubsection{Large radii}
\label{sec:soliton-large}

At large radii above a transition radius $r_{\rm sg}$ given by 
\begin{equation}
r_{\rm sg} = r_s \frac{\rho_a}{2 \rho_0} \gg r_s ,
\label{eq:rsg-def}
\end{equation}
the DM self-gravity dominates over the BH gravity and we recover the
soliton, or "boson star", profile. This is the spherically symmetric hydrostatic equilibrium 
of the Schr\"odinger-Poisson system. For a quartic self-interaction, in the Thomas-Fermi limit
(\ref{eq:large-mass}) where the self-interaction dominates over the "quantum pressure",
one obtains the density profile \cite{Brax:2019fzb},
\begin{equation}
\rho(r) = \rho_0 \frac{\sin(\pi r/R_{\rm sol})}{\pi r/R_{\rm sol}} , \;\;\; \mbox{with} \;\;\;
R_{\rm sol} = \pi r_a = \sqrt{\frac{\pi}{4 {\cal G} \rho_a}} .
\label{eq:soliton}
\end{equation}
This gives a finite-size soliton of radius $R_{\rm sol}$ and bulk density $\rho_0$.
Then, beyond the transition radius $r_{\rm sg}$ of Eq.(\ref{eq:rsg-def})
and below the soliton radius $R_{\rm sol}$ we have
\begin{equation}
r_{\rm sg} \ll r \ll R_{\rm sol} : \;\; \rho \simeq \rho_0 , \;\;\; 
v_r \sim - \frac{\rho_0}{\rho_a} \frac{r_{\rm sg}^2}{r^2} .
\label{eq:rho-vr-far}
\end{equation}
The DM density is almost constant whereas the radial velocity 
becomes negligible and falls off as $1/r^2$ as we converge to the static soliton.
The transition radius $r_{\rm sg}$ of Eq.(\ref{eq:rsg-def}) is much greater than the
BH horizon when $\rho_0 \ll \rho_a$ (i.e., we assume that the bulk of the dark matter cloud 
is nonrelativistic).
The effective pressure $P$ of the dark matter fluid, associated with the self-interaction, reads
\begin{equation}
P \simeq \frac{\rho_0^2}{2 \rho_a} = \frac{1}{2} \rho_0 c_s^2 , \;\;\; c_s^2 = \frac{\rho}{\rho_a} ,
\label{eq:P-large-r}
\end{equation}
where we introduced the sound speed $c_s$.

\subsubsection{Small radii}
\label{sec:soliton-small}

Below the transition radius (\ref{eq:rsg-def}), the BH gravity dominates and
the DM falls onto the BH increasingly fast. This leads to a DM density spike \cite{Brax:2019npi},
\begin{align}
r_s \ll r \ll r_{\rm sg}  : & \;\;\; \rho \simeq \rho_a \frac{r_s}{2 r} , \;\;\; 
P \simeq \frac{\rho^2}{2 \rho_a} \left[ 1 + 9 F_\star^2 \frac{r_s}{r} \right] , \nonumber \\
& \;\;\; v_r \sim - \frac{r_s}{r} .
\label{eq:rho-vr-close}
\end{align}
The second term in the expression of the pressure is the contribution from the radial inflow,
which becomes negligible for $r \gtrsim r_s$ as the flow becomes subsonic.
These expressions connect with the large-radii expressions (\ref {eq:rho-vr-far})
and (\ref{eq:P-large-r}) at the transition radius $r_{\rm sg}$ of Eq.(\ref{eq:rsg-def}).
Close to the BH the radial velocity is relativistic while the density is of the order of the characteristic
density $\rho_a$ introduced in Eq.(\ref{eq:rho-a-def}).

\subsection{Variety of scalar DM profiles}

\subsubsection{Steady-state pressure-supported profile}
\label{sec:steady-pressure}

Summarizing the results of the previous sections, we approximate the dark matter density profile
of the steady-state solution (\ref{eq:Fc}) as
\begin{eqnarray}
&& r_s < r < r_{\rm sg} : \;\; \rho = \rho_a \frac{r_s}{2 r} , \;\;\; 
r_{\rm sg} < r < R_{\rm sol} : \;\; \rho = \rho_0 , \nonumber \\
&& r > R_{\rm sol} : \;\;\; \rho = 0 ,
\label{eq:density-profile}
\end{eqnarray}
and we have $P=\rho^2/(2\rho_a)$ throughout.
It turns out that this density profile defines a critical soliton
mass $M_c$ associated with the case $R_{\rm sol}=r_{\rm sg}$, that is, the
$1/r$ density spike extends up to the radius $R_{\rm sol}$ of the DM halo.
This gives
\be
M_c = \pi \rho_a r_s R_{\rm sol}^2 = \frac{\pi^2}{2} M_0 \simeq 2 \times 10^7 \, M_\odot ,
\label{eq:Mc}
\ee
which is always of the same order as the BH mass and does not depend on the scalar field parameters.
This is because the characteristic density $\rho_a$ and radius $R_{\rm sol}$ are related by 
Eq.(\ref{eq:soliton}) and they cancel out in the expression (\ref{eq:Mc}) of $M_{\rm sol}$.
This means that if the DM cloud has an initial mass that is below $M_c$ it can never reach 
the Bondi-like pressure-regulated infall over the full radius $R_{\rm sol}$.
Thus, the full profile (\ref{eq:density-profile}) can only apply to the DM clouds that have a mass
greater than $M_c$.

\subsubsection{Free-falling DM cloud}
\label{sec:free-fall}

If there is no central BH, a soliton of radius $R_{\rm sol}$ can follow the hydrostatic profile
(\ref{eq:soliton}) for any value of the central density $\rho_0$, hence the soliton mass
$M_{\rm sol} \simeq (4 \pi/3) \rho_0 R_{\rm sol}^3$.
However, if a supermassive BH is embedded in this DM cloud, we have seen in (\ref{eq:Mc})
that low mass clouds below $M_c$ cannot reach the full steady-state solution (\ref{eq:Fc}).
If the initial DM density $\rho_0$ is too low, the pressure $P \propto \rho^2$ is too low to
slow down the infall onto the BH.
This corresponds to a fully supersonic free-fall solution, which reads in steady state as
\be
v_r = - \sqrt{\frac{2 {\cal G} M_0}{r}} , \;\;\; \rho = \frac{F}{4\pi r^2 |v_r|} ,
\ee
where the flux $F$ is below the critical flux $F_c$ of Eq.(\ref{eq:Fc}).
The lifetime $t_{\rm sol}$ of such a cloud of radius $R_{\rm sol}$ is given by
\be
t_{\rm sol} \sim \frac{R_{\rm sol}}{|v_r(R_{\rm sol})|} .
\ee
Requiring this lifetime to be greater than the age of the Universe $t_H$ gives
\be
t_{\rm sol} > t_H : \;\;\; R_{\rm sol} > r_s^{1/3} t_H^{2/3} \simeq 16 \, {\rm kpc} .
\label{eq:tsol-free-fall}
\ee
DM solitons should not have radii greater than 1 kpc to fit inside observed galaxy density profiles.
Therefore, realistic free-falling DM clouds should have already been eaten by the supermassive BH.

\subsubsection{Truncated pressure-supported profile $R_{\rm trunc} < R_{\rm sol}$}
\label{sec:Truncated}

If the initial DM density is large enough but the DM soliton is not very massive, depending on the
initial dynamics it may reach the transonic profile (\ref{eq:rho-vr-close}) truncated to a radius
$R_{\rm trunc} < R_{\rm sol}$, so that the total DM mass 
\be
M_{\rm trunc} = \pi \rho_a r_s R_{\rm trunc}^2 
\ee
is below the characteristic mass $M_c$ of Eq.(\ref{eq:Mc}).
The lifetime of such a cloud is 
\be
t_{\rm trunc} \sim \frac{R_{\rm trunc}}{|v_r(R_{\rm trunc})|} \sim \frac{R_{\rm trunc}^2}{r_s} ,
\ee
and requiring $t_{\rm trunc} > t_H$ gives the lower bound
\be
R_{\rm trunc} > \sqrt{r_s t_H} \simeq 35 \, {\rm pc} .
\label{eq:Rtrunc-bound}
\ee
This implies that the cloud radii can be much smaller than galactic sizes but are still much greater than the
dynamical radius (\ref{eq:Rdyn-SgrA*}).

\subsubsection{Outer virialized DM halo}
\label{sec:halo-outer}

As found in numerical simulations, DM halos that form from stochastic initial
conditions through gravitational instability, as on cosmological scales through hierarchical
structure formation, are not fully absorbed by a central soliton.
One typically obtains instead a central static soliton, as in (\ref{eq:soliton}),
embedded within an extended halo that follows a NFW profile like for CDM.
The quantum pressure and the self-interactions are negligible on this larger scale
and this extended halo is instead supported by the velocity dispersion of wave packets,
which play the role of the velocity dispersion of CDM particles \cite{Garcia:2023abs}.
In terms of the Schr\"odinger equation obeyed by the scalar field, where the
potential is dominated by the DM gravitational potential, these correspond to excited
energy eigenstate whereas the soliton corresponds to the ground state.
Although the soliton has a large occupation number and contains a macroscopic mass,
one often finds that a large fraction of the DM can remain in this extended halo.

\section{Impact of the dark matter environment on the shadow radius}
\label{sec:impact-steady}

\subsection{Low-mass DM soliton}

If the DM soliton has a mass below the critical mass (\ref{eq:Mc}), we have seen that it should
be a free-falling cloud as discussed in Sec.~\ref{sec:free-fall} or a truncated profile
as discussed in Sec.~\ref{sec:Truncated}.
As noticed below Eq.(\ref{eq:tsol-free-fall}), free-falling clouds should have already been eaten
by the supermassive BH. Therefore, we do not consider them further.
Truncated clouds can survive until today if their radius is above the lower bound (\ref{eq:Rtrunc-bound}).
Therefore, we have the hierarchy $r_s \ll R_{\rm dyn} \ll R_{\rm trunc}$.
Then, we obtain from Eq.(\ref{eq:Delta-bcrit})
\be
\frac{\Delta b_{\rm cr}}{b_{\rm cr0}} \simeq \frac{M_{\rm trunc}}{M_0} \left( \frac{r_s}{R_{\rm trunc}} - \frac{R_{\rm dyn}^2}{R_{\rm trunc}^2} \right) ,
\ee
where the mass of the truncated DM cloud is $M_{\rm trunc} = \pi \rho_a r_s R_{\rm trunc}^2$.
Using the upper bound (\ref{eq:Mc}) for the cloud mass and the lower bound (\ref{eq:Rtrunc-bound})
for the cloud radius, we obtain
\be
\left| \frac{\Delta b_{\rm cr}}{b_{\rm cr0}} \right| \lesssim 10^{-7} .
\ee
Thus, the impact on the shadow radius of such low mass and extended DM halos is completely negligible.

\subsection{High-mass DM soliton}

\subsubsection{Long-lived DM solitons}

We now consider massive DM solitons, above the critical mass (\ref{eq:Mc}). These clouds follow
the full profile (\ref{eq:density-profile}), with a constant density plateau $\rho_0$ at radii
$r_{\rm sg} < r < R_{\rm sol}$.
From Eq.(\ref{eq:rho-vr-far}) the lifetime of this soliton is
\be
t_{\rm sol} \sim \frac{12 {\cal G} M_{\rm sol} R_{\rm sol}^2}{\pi^2 r_s^2} ,
\ee
and requiring $t_{\rm sol} > t_H$ gives
\be
t_{\rm sol} > t_H : \;\;\; M_{\rm sol} R_{\rm sol}^2 > \frac{\pi^2 r_s^2 t_H}{12 {\cal G}} \simeq
10^{10} \, M_\odot \, {\rm pc}^2 .
\label{eq:lifetime-soliton}
\ee
At the lower mass bound, $M_{\rm sol}=M_c$, this gives $R_{\rm sol} \gtrsim 20 \, {\rm pc}$, 
in agreement with (\ref{eq:Rtrunc-bound}).
More massive clouds can have a smaller radius and still survive until today.

The mass of the Milky Way is estimated at about $M_{\rm MW} \simeq 2 \times 10^{11} M_\odot$.
We can expect the soliton mass to be a small fraction of the total galaxy mass, 
therefore we take the rather conservative upper bound on the DM soliton mass
\be
M_{\rm sol} < M_{\rm max} \;\;\; \mbox{with} \;\;\; M_{\rm max} = 0.1 M_{\rm MW} \simeq 
2 \times 10^{10} M_\odot ,
\label{eq:Mmax}
\ee
With Eq.(\ref{eq:lifetime-soliton}), this gives a lower bound on the radius $R_{\rm sol}$ for the 
soliton to survive until today,
\be
R_{\rm sol} > R_{\rm min} \;\; \mbox{with} \;\;
R_{\rm min}= \sqrt{ \frac{\pi^2 r_s^2 t_H}{12 {\cal G} M_{\rm max}} } \simeq 0.6 \, {\rm pc} .
\label{eq:Rmin}
\ee
Therefore, the stellar orbital radius $R_{\rm dyn}$ of Eq.(\ref{eq:Rdyn-SgrA*}) is much smaller
than $R_{\rm sol}$ and we have the hierarchies $r_s \ll r_{\rm sg} < R_{\rm sol}$ and
$r_s \ll R_{\rm dyn} \ll R_{\rm sol}$.
Then, we have two cases depending on whether $R_{\rm dyn}$ is below or above the transition
radius $r_{\rm sg}$, which from Eq.(\ref{eq:rsg-def}) also reads as
\be
r_{\rm sg} = \frac{\pi^2 M_0}{3 M_{\rm sol}} R_{\rm sol}.
\label{eq:rsg-Msol}
\ee

\subsubsection{$r_s \ll R_{\rm dyn} < r_{\rm sg} < R_{\rm sol}$}

We first consider the case $R_{\rm dyn} < r_{\rm sg}$, which from Eq.(\ref{eq:rsg-Msol}) gives
\be
R_{\rm dyn} < r_{\rm sg}  : \;\;\; R_{\rm sol} > \frac{3 M_{\rm sol}}{\pi^2 M_0} R_{\rm dyn} .
\label{eq:large-rsg}
\ee
Then, Eq.(\ref{eq:Delta-bcrit}) gives
\be
\frac{\Delta b_{\rm cr}}{b_{\rm cr0}} \simeq \frac{M_{\rm sol}}{M_0}  
\left( \frac{3 r_s}{4 R_{\rm sol}} - \frac{3 r_{\rm sg} R_{\rm dyn}^2}{2 R_{\rm sol}^3}
\right) .
\label{eq:Delta-bcr-high-mass-1}
\ee
This is bounded by
\be
\left| \frac{\Delta b_{\rm cr}}{b_{\rm cr0}} \right| \leq \frac{3 M_{\rm max} r_s}{4 M_0 R_{\rm min}}
+ \frac{\pi^2 R_{\rm dyn}^2}{2 R_{\rm min}^2} \simeq 0.002 + 0.0003 ,
\ee
where we used Eq.(\ref{eq:rsg-Msol}) in the second term.
This is always much below the observational bound (\ref{eq:observ}).

\subsubsection{$r_s \ll r_{\rm sg} < R_{\rm dyn} \ll < R_{\rm sol}$}

We now consider the case $R_{\rm dyn} > r_{\rm sg}$, which from Eq.(\ref{eq:rsg-Msol}) gives
\be
R_{\rm dyn} > r_{\rm sg}  : \;\;\; R_{\rm sol} < \frac{3 M_{\rm sol}}{\pi^2 M_0} R_{\rm dyn} .
\label{eq:small-rsg}
\ee
Then, Eq.(\ref{eq:Delta-bcrit}) gives
\be
\frac{\Delta b_{\rm cr}}{b_{\rm cr0}} \simeq \frac{M_{\rm sol}}{M_0} 
\left( \frac{3 r_s}{4 R_{\rm sol}} - \frac{R_{\rm dyn}^3}{R_{\rm sol}^3} \right) .
\ee
This is bounded by
\be
\left| \frac{\Delta b_{\rm cr}}{b_{\rm cr0}} \right| \leq \frac{3 M_{\rm max} r_s}{4 M_0 R_{\rm min}}
+ \frac{M_{\rm max} R_{\rm dyn}^3}{M_0 R_{\rm min}^3} \simeq 0.002 + 0.002 ,
\ee
which is again much below the observational bound (\ref{eq:observ}).

\subsection{Short-lived solitons}
\label{sec:short-lived}

We have seen in the previous sections that long-lived central solitons cannot be constrained by the
measurement (\ref{eq:observ}) of the shadow radius, as clouds that are large enough to have a long
lifetime are also too diffuse to make a significant impact on the shadow radius.
However, as recalled in Section~\ref{sec:halo-outer}, numerical simulations show that 
solitons are not isolated compact DM clouds but are embedded within extended halos 
that behave like CDM. Within the cosmological context, we expect such a virialized DM halo
to follow the mean NFW density profile as for CDM scenarios and to extend beyond
50 kpc, forming most of the mass of the Milky Way.
Then, we can imagine that a short-lived soliton can exist if it is continuously replenished by the
outer DM halo while it loses mass to the supermassive BH.
However, if this is the case, its mass must be much smaller than the mass $M_0$ of the central BH,
since by the same assumption the BH has already "eaten" several generations of the soliton.
Then, the impact on the shadow radius is small, since it is clear from Eq.(\ref{eq:Delta-bcrit}) that
we have the generic upper bound
\be
\left| \frac{\Delta b_{\rm cr}}{b_{\rm cr0}} \right| \lesssim \frac{M_{\rm sol}}{M_0} .
\label{eq:Deltab-Msol}
\ee
This means that solitons with a lifetime below $t_H/10$ should also have $M_{\rm sol}<M_0/10$
and $| \Delta b_{\rm cr} /  b_{\rm cr0} | \lesssim 0.1$. Therefore, they automatically satisfy the
observational upper bound (\ref{eq:observ}).

\subsection{Outer DM halo}
\label{sec:outer-DM-halo}

The extended virialized DM halo, which is supported by its velocity dispersion instead of self-interactions 
or quantum pressure, has a very long lifetime and a large mass. 
Therefore, we briefly consider the constraints on this DM halo that can be derived from
the measurement (\ref{eq:observ}) of the shadow radius.
For a generic treatment, we assume a power-law dark matter density profile,
\be
r_s \ll r < R_{\rm halo} : \;\; \rho = \rho_0 \left( \frac{r}{r_s} \right)^{-\alpha} ,
\ee
with
\be
\rho_0 \ll \rho_a , \;\;\; 1 \leq \alpha < 3 , \;\;\;
M_{\rm halo} \simeq \frac{4\pi\rho_0}{3-\alpha} R_{\rm halo}^3 
\left( \frac{R_{\rm halo}}{r_s} \right)^{-\alpha} ,
\ee
and we take the pressure to be negligible (nonrelativistic dust).
The halo mass is dominated by the outer shells.

If the DM halo radius $R_{\rm halo}$ is smaller than the stellar radius $R_{\rm dyn}$
used to estimate the central dynamical mass, we obtain
$\frac{\Delta b_{\rm cr}}{b_{\rm cr0}} \simeq - \frac{M_{\rm halo}}{M_0}$.
Then the measurement of the shadow radius implies that the DM halo mass is smaller than the
BH mass. However, this is not realistic, as the virialized DM halo extends much beyond $R_{\rm dyn}$
and is more massive than the central BH.
Therefore, we have $R_{\rm dyn} \ll R_{\rm halo}$ and we obtain
\be
\frac{\Delta b_{\rm cr}}{b_{\rm cr0}} \simeq \frac{M_{\rm halo}}{M_0} 
\left[ - \left( \frac{R_{\rm dyn}}{R_{\rm halo}} \right)^{3-\alpha} + \frac{3-\alpha}{2(2-\alpha)} \frac{r_s}{R_{\rm halo}} \right] .
\ee
With $R_{\rm halo} \geq 10 \, {\rm kpc}$ and $M_{\rm halo} \simeq 2 \times 10^{11} M_\odot$, 
the second term gives a contribution below $10^{-5}$ that is much below the observational upper 
bound (\ref{eq:observ}).
The first term gives a contribution below $10^{-7}$ for $\alpha=1$ and below $0.025$ for $\alpha=2$.
These conservative results show that realistic DM halos satisfy the observational constraint 
(\ref{eq:observ}).

\section{Discussion}\label{sec:discussion}

Adopting a phenomenological approach, we have investigated whether the observations of Sgr A$^{\star}$ 
from the EHT collaboration can provide competitive constraints on DM models, focusing on the case
of scalar DM. Firstly, we have presented a generic perturbative scheme to examine the influence
of the environment (baryon or DM clouds) on the metric and, consequently, on the BH shadow radius. 
Secondly, in the case of scalar DM with non-negligible self-interactions, we have recalled the 
properties of the Bondi-like accretion flow onto the supermassive BH and its dependence on the
BH mass and the scalar-field parameters. This applies to the large-mass limit, $m\gg 10^{-17}$~eV,
where the Compton wavelength is smaller than the Schwarzschild radius.
This accretion flow describes the loss of matter from the central soliton -- the ground state 
configuration (of the Schr\"odinger equation of motion that governs the DM dynamics) that forms 
in the galactic potential well -- to the central supermassive BH.
Thirdly, we examined the impact of such a soliton onto the shadow radius, as well as that of
the outer virialized DM halo that extends beyond the galactic radius as in standard CDM scenarios.

We find that physical self-consistency conditions of the underlying DM model strongly constrain
the impact of the DM environment on the observed shadow radius.
First, we noted that a full Bondi-like steady-state accretion flow can only be realized for DM solitons
that have a mass which is greater than that of the supermassive BH. Smaller-mass solitons either fall
onto the BH as in free fall (the DM density is too low for the effective pressure to slow down the
infall) and have a lifetime that is much smaller than the Hubble time, or exhibit a truncated profile
(they are throughout dominated by the BH gravity). In this latter case, requiring that their lifetime
is greater than the Hubble time implies a negligible impact on the shadow radius (because
a long lifetime implies a large cloud radius whence a small DM central density).

We next found that solitons that are more massive than the central BH (and thus show a complete
Bondi-like transonic accretion flow, which goes from free fall near the horizon to hydrostatic
equilibrium at large radii, dominated by the self-gravity of the DM cloud) have a small impact
on the shadow radius, somewhat below the observational upper bound (\ref{eq:observ}).

Taking into account the outer DM halo, which should resemble the NFW profile found in CDM scenarios,
we find that if it can replenish short-lived solitons the latter should have a small mass and
a small impact on the shadow radius, while the impact of this virialized DM halo itself should also
be small.

Thus, our results show the importance of taking into account the self-consistency requirements
of underlying DM models. Within a given framework, it is not possible to assume arbitrary DM density 
profiles. Taking into account the accretion flow onto the BH and the lifetime of peculiar configurations
significantly limits the range of possible DM profiles and the impact on the observed shadow radius.
In the specific case of scalar field DM, which can form a high-density soliton at the center of the
galactic gravitational potential well, within an extend NFW-like virialized halo, we have shown that 
such constraints are sufficient to ensure that the impact of the DM environment on the 
shadow radius remains well within the observational bounds.

\section*{Acknowledgments}

G. G. acknowledges financial support from Agencia Nacional de Investigaci\'on y Desarrollo (ANID) through the FONDECYT postdoctoral Grant No. 3210417.

\bibliography{biblio}

\end{document}